\documentclass[12pt,preprint]{aastex}

\shorttitle{Galactic magnetic halo and UHECR}
\shortauthors{Chardonnet \& Mattei}

\begin{document}

 \title{On the role of galactic magnetic halo in the ultra high energy cosmic rays propagation.}

\author{Pascal Chardonnet\altaffilmark{1} and Alvise Mattei\altaffilmark{2}}
\affil{LAPTH, BP 110 F-74941, Annecy-le-Vieux Cedex, France}

\altaffiltext{1}{Universit\'e de Savoie; email: chardonnet@lapp.in2p3.fr}
\altaffiltext{2}{mattei@lapp.in2p3.fr }

\begin{abstract}

The study of propagation of Ultra High Energy Cosmic Rays (UHECR) is a key step in order to unveil the secret of their origin. Up to now it was considered only the influence of the galactic and the extragalactic magnetic fields. In this article we focus our analysis on the influence of the magnetic field of the galaxies standing between possible UHECR sources and us. Our main approach is to start from the well known galaxy distribution up to 120~Mpc. We use the most complete galaxy catalog: the LEDA catalog. Inside a sphere of 120~Mpc around us, we extract 60130 galaxies with known position. In our simulations we assign a Halo Dipole magnetic Field (HDF) to each galaxy. The code developed is able to retro-propagate a charged particle from the arrival points of UHECR data across our galaxies sample. We present simulations in case of Virgo cluster and show that there is a non negligible deviation in the case of protons of $7 \times 10^{19}$ eV, even if the $B$ value is conservative. Then special attention is devoted to the AGASA triplet where we find that NGC3998 and NGC3992 could be possible candidates as sources.

\end{abstract}


\notetoeditor{Figures f3.eps and f4.eps should appear one above the other, with only one caption}

\section{Introduction: two puzzles connected}

The window of observational astronomy has become broader, firstly in the '70s with the development of the first gravitational wave detectors, then in the '90s with the birth of the astroparticle field and the construction of neutrinos telescope. On the other hand, the charged particles, produced abundantly in all astrophysical sources, lose the direction of the source, due to diffusion inside galactic magnetic field (GMF). However, one hope is to use the highest part of the cosmic ray spectrum, to open a new window in astrophysics. To do this, we must find a path connecting the two fundamental ingredients: the ultra high energy cosmic rays (UHECR) and the extra-galactic magnetic field (EGMF). The existence of a $50~\mu$G GMF was discovered by \cite{Biermann}. EGMF of $\sim \mu$G were detected by both synchrotron emission and Faraday rotation in the Coma Cluster. Beyond clusters, in sparser regions, surprising magnetic fields of $\sim 0.1~\mu$G were measured \citep{Kim}. Observations suggest that magnetic fields are present in all galaxies and clusters. Even if they are small, they are extended on very large scale and the energy inside could be very huge. More details are in \cite{Han}, \cite{Beck} and \cite{Vallee}.

The importance of magnetic field is well known in the case of radio galaxies where the radio lobes confirm the existence of a large magnetized region well outside the galaxy. \cite{Kronberg} advanced the idea that the GMF, especially in the case of radio galaxies, is a consequence of the presence of a supermassive black hole at the center. Following this, and using the fact that most probably all galaxies have a massive black hole in their center, we will assume that the existence of the galactic magnetic halo is a general property of galaxies. This has to be modulated by the value of the related field and the size of this halo. There is also evidence for the existence of the magnetic halo in the case of spiral galaxy seen edge-on \citep{krause}.

On the other hand, the problem of the UHECR origin is a long standing mystery, starting with the discovery of a $10^{20}$~eV event \citep{Linsley} that opened the Pandora's box. More details can be found in \citet{Olinto1}. Magnetic fields play a vital role in this puzzle, and are a corner stone for the UHECR propagation. The gyroradius $R$ of a proton of energy $E$ in a magnetic field $B$ is given by $R\sim \mathrm{100~kpc} \times E(10^{20}\mathrm{eV}) /B(\mu\mathrm{G})$. The influence of cosmic rays propagation has been underlined in several articles. \cite{Alvare} scrutinized the influence of GMF and they showed that protons in the range of $10^{18} \div 10^{19}$~eV can probe the large scale structure of GMF. \cite{Prouza} made a study of the influence of the GMF, including a magnetic galactic halo in form of dipole given by \cite{Han}. \cite{Tanco} analyzed the effect of large EGMF localized in sheets and filaments, while voids are kept almost free of magnetic field. He showed that this assumption has dramatic consequences on the cosmic ray propagation. \cite{Sig} analyzed the influence of EGMF and showed that for strongly magnetized regions, observers predict considerable large scale anisotropies between $10^{19}$~eV and $10^{20}$~eV. \cite{yoshi} presented numerical UHECR simulations in a structured intergalactic magnetic field given by IRAS PSCz catalog. They estimated the number of sources to explain the AGASA data observations and found that $\sim 10^{-5}\div 10^{-6}$~Mpc$^{-3}$ is the best number of UHECR sources density. \cite{Dolag} examined the UHECR propagation using simulations of large-scale structure formation to study the build-up magnetic field in the intergalactic magnetic medium. They constructed full-sky maps of expected UHE protons deflections. Cosmic rays propagation in the galaxy is strongly affected by the GMF, where the gyroradius of a proton can be as small as $\sim 300$~pc. Therefore, for $E<10^{18}$ eV, cosmic rays diffuse in the GMF. They get isotropized and hence do not reveal the sources where they were accelerated. In the range $10^{18} \div 10^{19}$ eV, cosmic rays through the magnetic fields are thought to change from heavy to light in composition and from diffusive to ballistic in propagation \citep{Aloisio}.

To study UHECR propagation we need the full map of the magnetic field. The source of this field relies on three different origins: the galactic one which is well known and for which we have an analytical expression; the extragalactic one, in main part completely unknown, and for which a lot of numerical simulations has been made using large scale structure formation; finally the effect of the field of the galaxies encountered by UHECR. It has long been hypothesized that the deviation by these galaxies are unlikely or inefficient. Since we have now a complete view of our local universe (up to 120~Mpc), thanks to many surveys, it is interesting to evaluate quantitatively this contribution. In this article we reconsider this hypothesis in light of a complete galaxy catalog. Lacking general data on magnetic halos for all the galaxies in the survey, we assume this is a general property and apply to each galaxy a model similar to the Milky Way.

In sec.~\ref{deux} we detail the way to reconstruct the nearby universe magnetic fields. In sec.~\ref{trois} we present the code used to retro-propagate the charged particles, here assumed to be protons, and the results of simulation in the Virgo cluster and maps of AGASA data, with particular attention to the triplet. In sec.~\ref{conc} we finally show that such deviations are not negligible and are to be incorporated in propagation calculations.

\section{The magnetic field map}
\label{deux}

To study UHECR propagation in the local universe, we looked at the most complete galaxy map available. Recently, \cite{Paturel} published a complete galaxy catalog by merging several catalogs. LEDA (Lyon-Meudon Extragalactic Database) is a collection of near infrared properties of galaxies, using full resolution images from DENIS measurement survey. This leads to a catalog of 753153 galaxies with I, J, L, K magnitudes. From this catalog, we extracted a sample of 60130 galaxies within 120~Mpc from us. We selected the galaxies for which we have a distance estimator in the catalog. LEDA permits also to have the inclination parameter of the galaxy axis as well as several other features.

We will use this sample to construct a magnetic map. Lacking direct measurement of magnetic field of large sample of galaxies, we will assume a simple model for it: each galaxy is surrounded by a magnetic halo with a dipole structure. The Halo Dipole magnetic Field (HDF) is, in spherical coordinates,
\begin{eqnarray*}
B_x = & - 3\mu_G \cos\phi \sin\phi \sin\theta / R^3 \\
B_y = &- 3\mu_G \cos\phi \sin\phi \cos\theta / R^3 \\
B_z = & \mu_G (1 - 3 \cos^2\phi)/ R^3
\end{eqnarray*}
where $\mu_G=184.2$~$\mu$G~kpc$^3$ is the galactic magnetic moment and $R$ is the distance from galactic center. These values are the same of Milky Way \citep{Stanev}. The inner zone ($< 500$~pc) is taken as a constant value $B_z = 500$~mG. The magnetic field is considered unimportant for $R > 100$~kpc, where it falls below $0.1$~nG and is set to zero. For this value, the Larmor radius of a $10^{20}$ eV proton is $\simeq  1$~Gpc. The orientation of the magnetic dipole $z$-axis is chosen with a random number, but is parallel to symmetry axis for elliptic galaxies or rotation axis for spiral ones.

According to \cite{Longair}, the average cosmic ray density in the Milky Way is $\sim 0.6$ eV~cm$^{-3}$. The total magnetic energy stored inside this halo is $(B^2/2\mu_0) (4\pi r_0^3/3) \sim10^{55}$~ergs, which is less than the total energy inside cosmic rays estimated as the value $10^{63}$~ergs for a normal galaxy. Therefore the $B$ value in this assumption is conservative and it is much less than the equipartition one, which is reasonable for structured fields. By comparison, under the assumption that the field was adiabatically compressed, \cite{Furlanetto} found that the intergalactic magnetic field should be at $\sim 1$~nG, while we are at $0.1$~nG at the halo edge. We computed the average value of the magnetic field inside a box of 40~Mpc centered on our galaxy. The numerical integration gives $\sqrt{\left<  B^2 \right> } \simeq 2$~nG. It is compatible with Faraday rotation measurements of high $z$ radio loud QSO \citep{Carilli}. By comparison, \cite{Sig} found an average value $\sqrt{\left<  B^2 \right> } \simeq 79 $~nG with coherence lengths $\lesssim$~Mpc in the strong field regions. To explain such magnetized halo, we follow the arguments proposed by \cite{Kronberg} regarding Giant Radio Galaxies (GRG). Analyzing GRG, they concluded that a large fraction of energy is stored in extended radio source in form of magnetic energy. They noticed that, contrary to the case of radiation energy, which is quickly lost, magnetic energy gets confined for a significant fraction of cosmic time. They assumed that gravitational energy lies at the origin of this magnetized energy and proposed a new link between massive black holes and the magnetized universe. Since we have now solid indications that massive black holes are a common feature of all galaxies, we can extend the argument developed by \cite{Kronberg} to all galaxies: at some level the energy exchanges between gravitational and magnetic energy, through the presence of a massive black hole. The natural conclusion is that massive black holes are at the origin of magnetized galaxies. In the case of AGN, the presence of magnetism is unveiled by the radio lobes. In case of less active galaxies, it could be associated to a less extended lobe, like a magnetized halo. Our assumption is then fully justified in this new scenario.

\section{Code description and results}
\label{trois}

\begin{figure}
\epsscale{.80}
\plotone{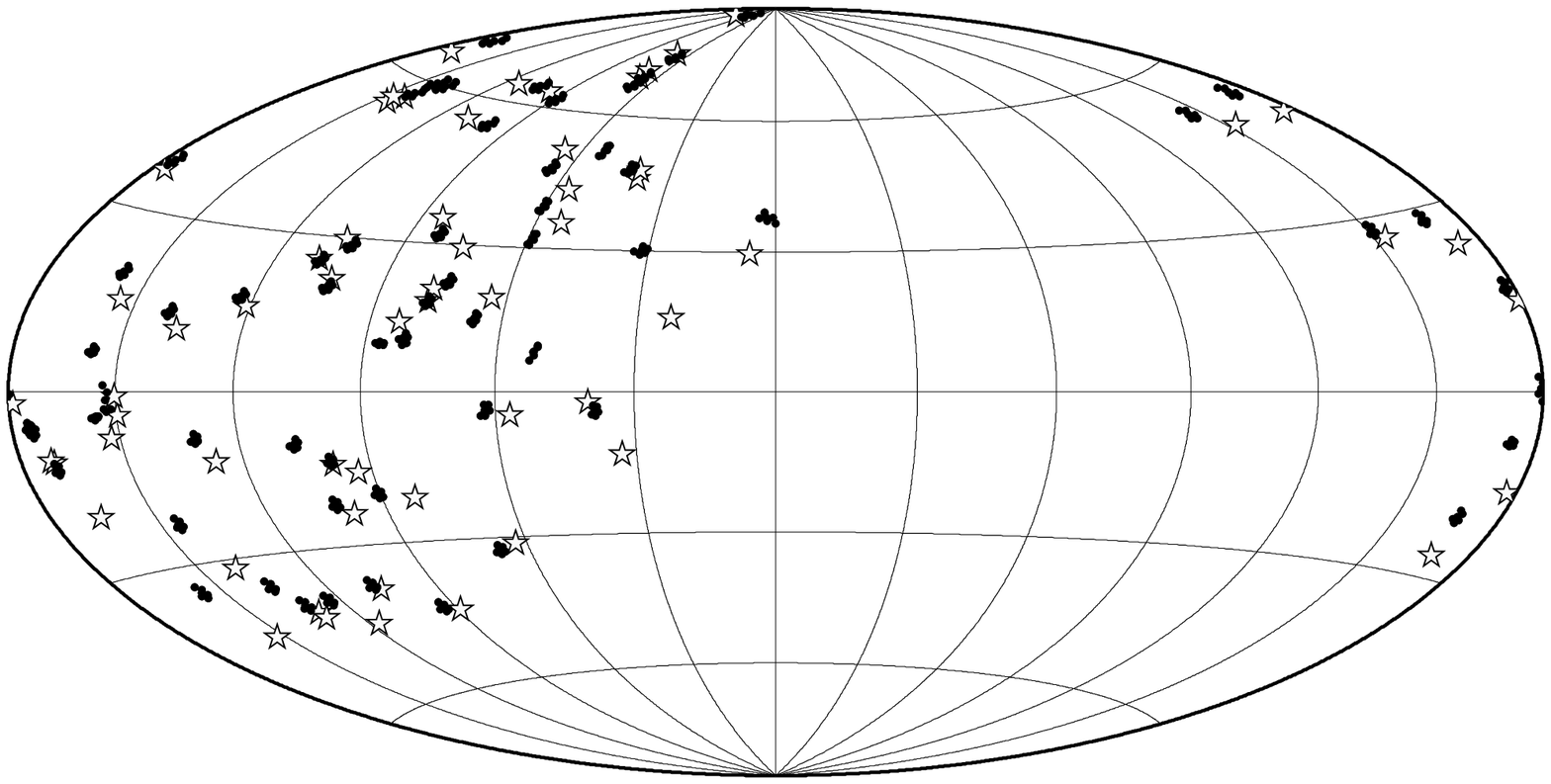}
\caption{Effect of the Milky Way magnetic field on AGASA data (stars). Hammer equal area projection in galactic coordinates. Each simulation has seven points inside a $1^\circ$ radius spot.}
\label{fig1}
\end{figure}

In our simulation we used the AGASA data \citep{agasa}. In fig.~\ref{fig1} we superimposed the AGASA events and the corresponding events after the interaction with GMF according to the BSS-A model given by \cite{Stanev}. The simulation shows clearly a GMF effect. The maximum deviation is related to region of high density of matter where $B$ is stronger. The results were checked for all the directions in the sky and were consistent with \cite{kst}.

\begin{figure}
\plotone{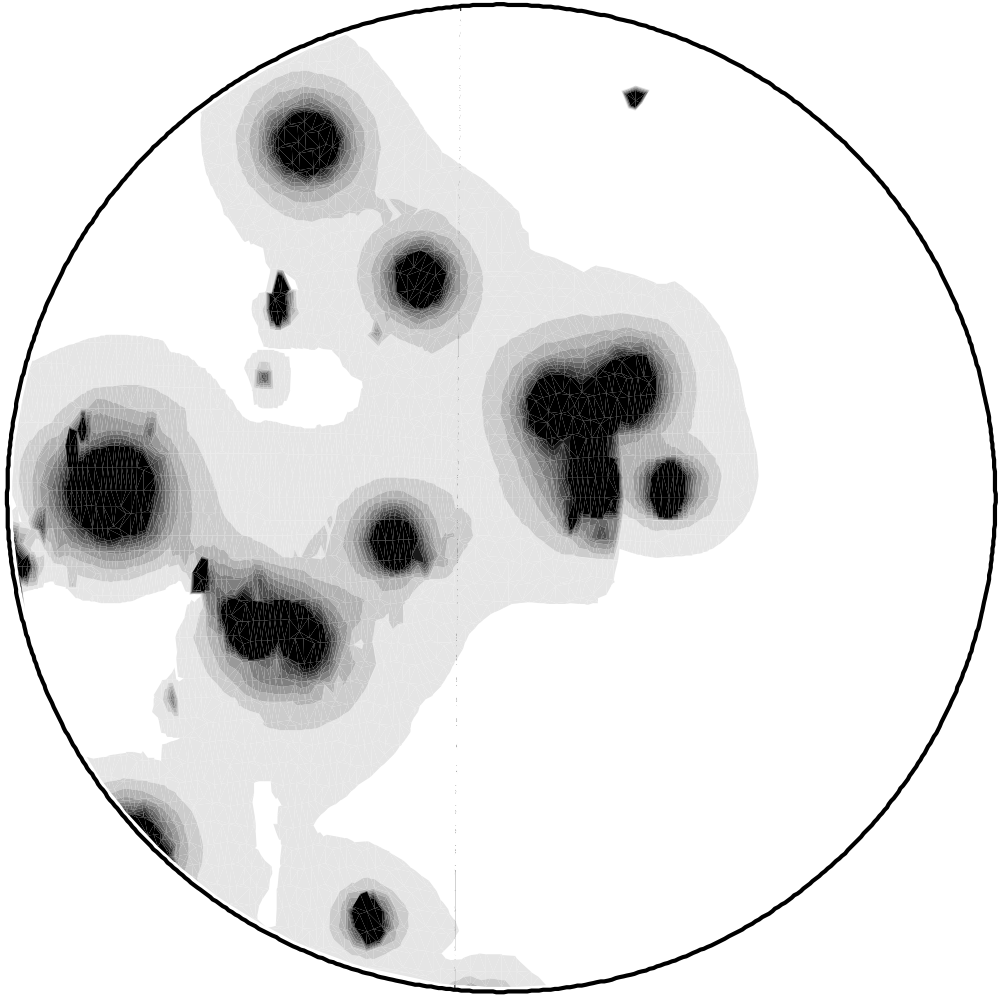}
\caption{Deflection for 10000 trajectories of energy $E=7\times10^{19}$~eV inside a circular spot of half radius 1$^\circ$ towards Virgo Cluster center, considered at 30~Mpc. We evidence on this picture that some part of the initial spot are deflected by angle greater than $1^\circ $ (Black zones) }. 
\label{fig2}
\end{figure}

Then we focused on two regions of the sky. The first is Virgo cluster where the mean galaxy concentration is twice the density background \citep{Char}. We inject 10000 monochromatic protons distributed in a cone half angle $1^{\circ}$ centered at $l=279.2$, $b=+74.4$, and study the backward propagation by numerical integration of the equation of motion. The integration step is variable, depending on magnetic field intensity, but $\leq 1$~kpc. The Lorentz equation depends on the ratio $ ZeB/ E$, so we adapted the steps as a function of the value of $\bf{B}(\vec{r})$ in order to conserve the accuracy of numerical solution. The magnetic field is neglected when $\leq0.1$~nG. Excluding the Milky Way contribution, the mean deflection due to HDF is $\Delta \theta \sim 0.71^\circ $ at 30~Mpc for a proton of energy $7\times10^{19}$. The result is given on fig.~\ref{fig2}. Adding the GMF contribution, we obtain a global deviation of $\Delta \theta \sim 1.56^\circ $. To evaluate the influence of the field amplitude, we made another simulation with a field frozen at 1~nG up to 300~kpc from the galaxy center (a \emph{frozen halo}). The results are compared in fig.~\ref{fig3ab}: even in the pure dipole approach we have some points at large angular deviation; while, for a larger magnetic field, the spot itself is spread by multiple deviations. The effect of possible deviations by galaxies between us and the source of UHECR could not be negligible. The second region corresponds to the AGASA triplet, whose center is located in the galactic coordinates ($l=142$; $b=+58$) where the galaxies density is low. Using our propagation code, with both GMF and HDF contribution, we launched 40000 trajectories for each of the triplet points, with the AGASA energy. We found that there is a small but important dispersion. Looking back to the trajectories we identified some common origin for the AGASA triplet (fig.~\ref{fig4}). Among the possible sources for the triplet we have isolated NGC3998 and NGC3992 which are at a distance of $\sim$ 25~Mpc. At least one trajectory for each triplet event passes through these galaxies.

NGC3998 is a LINER type I and it has been recently observed by XMM-Newton \citep{Ptak}. Several observations show an increase of X-ray flux, characteristic of an intense activity. They concluded that a jet model may be consistent with the radio and X-ray flux and jet emission may produce also the mid-IR and optical/UV flux. If we follow this conclusion, we can assume the existence of a jet and presumably a shock wave at the end of this jet. In this environment we expect to have UHECR acceleration \citep{rb}. We naturally propose NGC3998 as a solid candidate to be the AGASA triplet source.

NGC3992 is a giant luminous spiral galaxy located inside a galaxy group with a large mass to light ratio. Historically this is a known galaxy (M101), and hosted a supernova type I \citep{zwi}. It is interesting in light of the observed connection between supernovae and gamma-ray bursts, which are likely transient UHECR sources \citep{vietri}.

\section{Conclusions}
\label{conc}

\begin{figure}
\epsscale{.50}
\plotone{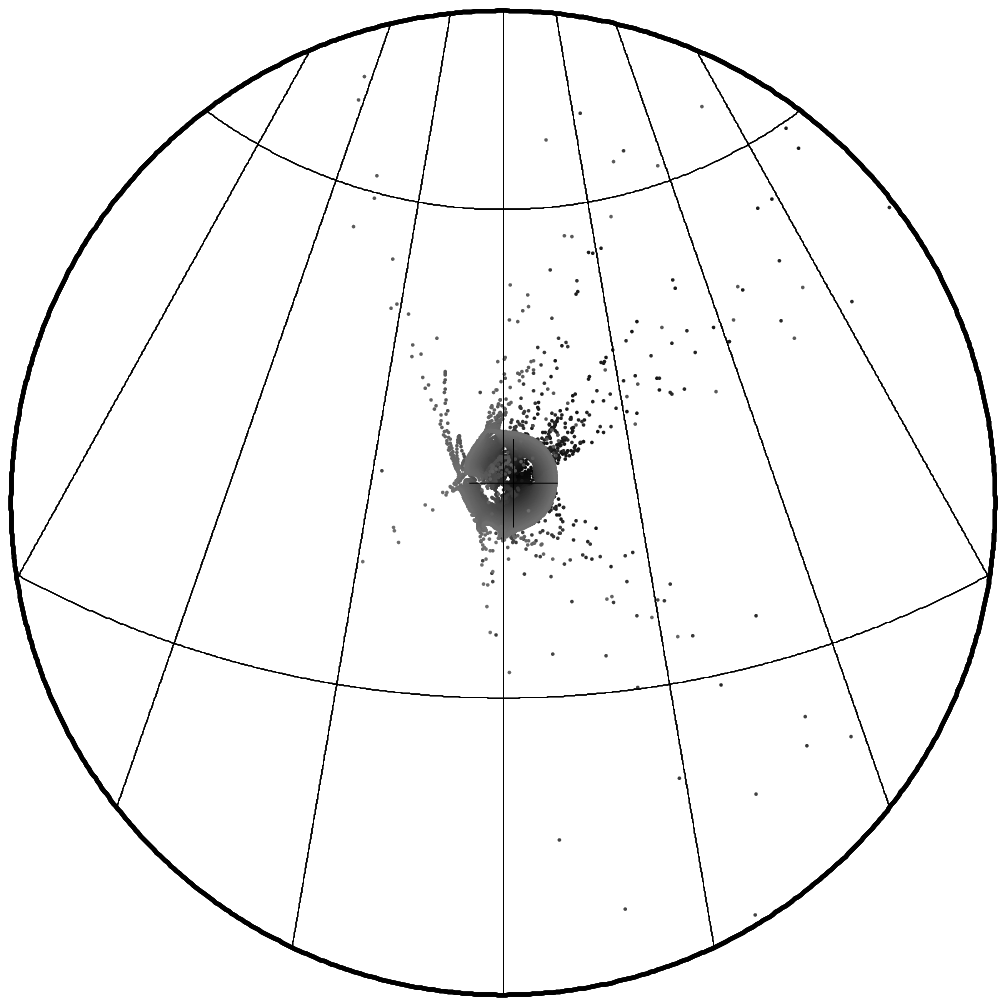}
\plotone{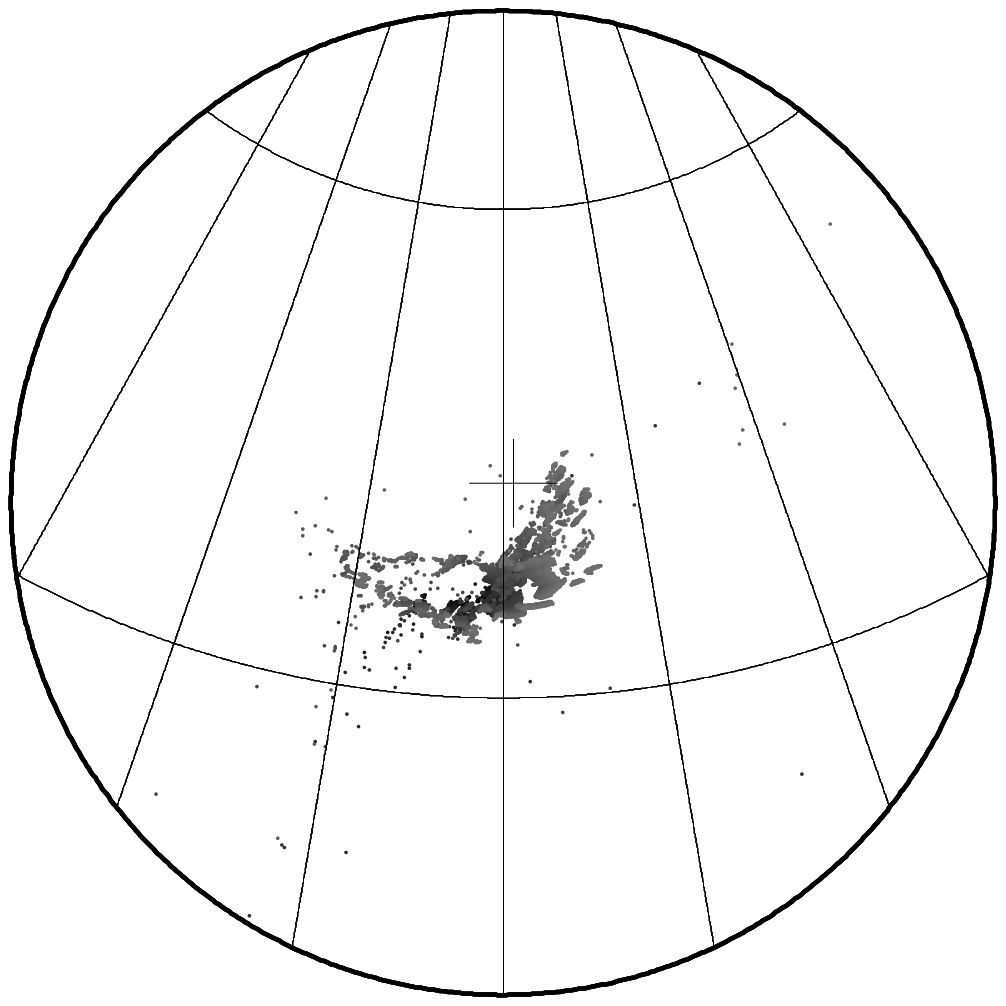}
\caption{The dispersion of trajectories through Virgo galaxies magnetic fields at 30~Mpc in gnomonic projection. 10000 protons at $E=7\times10^{19}$~eV, in a spot of half angle $1^\circ$ are injected. At start, darker points were closer to center of Virgo nominal coordinates ($l=279.2$, $b=+74.4$). Upper: a simple dipole field, as described in text; the mean deviation is $\Delta\theta=0.71^\circ$. Lower: the same, with a radial dipole magnetic field component frozen to $1$~nG up to a radius of $\sim 300$~kpc; the mean deviation is $\Delta\theta=2.03^\circ $. The upper picture shows that even in the conservative approach we have some points at large angular deviation. In the lower picture, the magnetic field is greater and the spot itself is highly deformed by multiple deviations.
\label{fig3ab}}
\end{figure}

\begin{figure}
\epsscale{.70}
\plotone{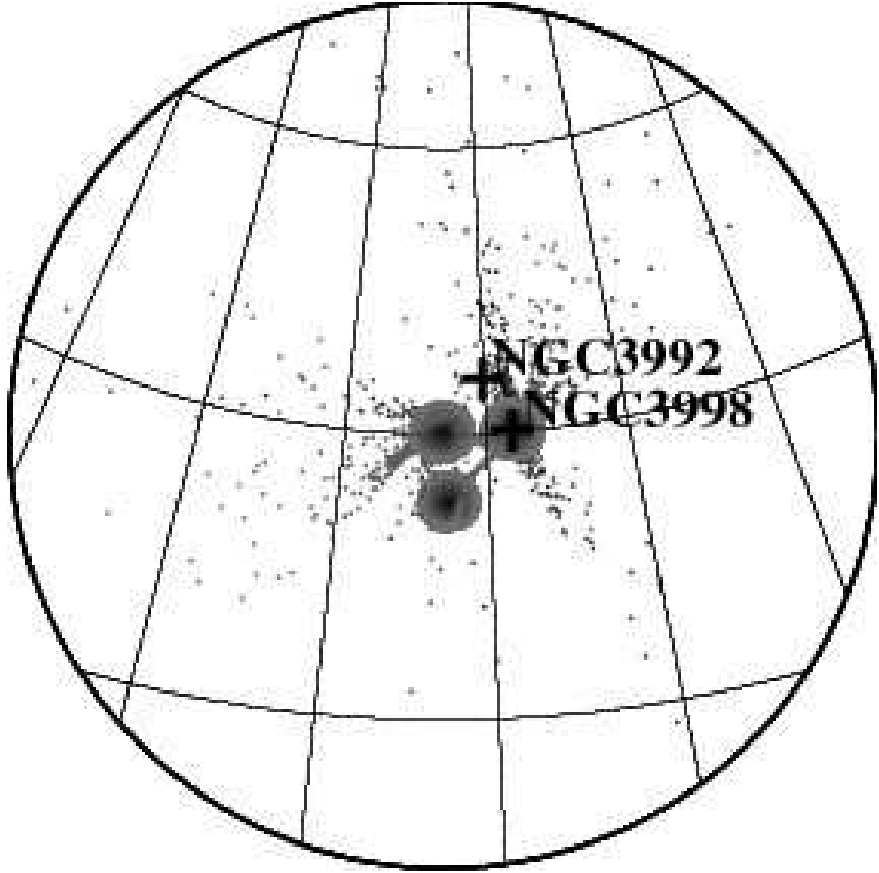}
\caption{Triplet dispersion at 30~Mpc. 40000 protons for each AGASA point (direction and energy) were back-propagated, including Milky Way field and outer galaxy dipole field. This triplet is located at galactic coordinate  l=142 ; b=+58. At start, darker points were closer to center of AGASA nominal coordinates. Two candidates are shown. To validate a source we choose the presence in at least one trajectory per event.
\label{fig4}}
\end{figure}

We developed a UHECR propagation code through the magnetic field constructed with the 60130 galaxies within 120~Mpc from LEDA catalog. In the simulations we assumed that UHECR are protons. Using AGASA data we found evidence for deviation that could affect the direction of the source. In the conservative approach the mean deviation we obtained is close to the one obtained by \citet{Dolag} by studying the cosmic rays propagation in the extragalactic magnetic field using the large scale structure (LSS) formation. Our result shows that the magnetic deviation by galaxies is not negligible. In fact, to do charged particle astronomy, we certainly should consider the granularity effect of our local universe and also the deviation associated. In that sense our work is a step further and complementary to the work using LSS simulations of \cite{Dolag}. The results are model dependent and our main objective is to demonstrate the influence of galactic magnetic halos on UHECR propagation. We want to underline that we used a conservative approach: first, in the choice of the B field we took a dipole component with an edge value of the halo of $ \sim 0.1$~nG; second, in our simulation we focused on protons at energy $ \sim 7\times10^{19} $~eV. Heavier or less energetic particles are more deviated. Despite all this, our work shows clearly that the influence of HDF is not negligible. A next step will be to incorporate in the code the influence of the cluster and the intergalactic magnetic fields, which has been partially done in fig.~\ref{fig3ab}. In the case of the AGASA triplet, our results suggest two candidates by tracing back the deflected trajectories. NGC3998 presents some astrophysical interest: it is an active galaxy at relatively short distance. Therefore we suggest it as a possible UHECR source. The scenario presented here is related to the galactic magnetic halo and answers to the fundamental question of the relation between magnetism and structure formation. This question is addressed from the theoretical part with some possible connection between supermassive black holes and magnetism, as in Kronberg scenario. The SKA and LOFAR survey will surely clarify the picture by mapping the magnetic fields with more accuracy. UHECR propagation could give us another astrophysical test of this connection.

We would like to thank G. Paturel and P. Prugniel for valuable discussions. We are also grateful to the CC-IN2P3 at Lyon for technical support in computing. We acknowledge the referee's contribution in the paper discussion.

\clearpage 

\clearpage 

\clearpage 

\clearpage 

\begin{thebibliography}{}

\bibitem[Alvarez-Muniz \& Stanev(2005)]{Alvare} Alvarez-Muniz, J., \& Stanev, T. 2005, preprint (astro-ph/0507273)
\bibitem[Aloisio \& Berezinsky(2005)]{Aloisio} Aloisio, R. \& Berezinsky, V. A. 2005, \apj, 625, 249
\bibitem[Beck(2004)]{Beck} Beck, R. 2004, in How Does the Galaxy Work?, ed. E. J. Alfaro et al. (Dordrecht: Kluwer), 277
\bibitem[Carilli \& Taylor(2002)]{Carilli} Carilli, C. L., \& Taylor, G. B. 2002, \araa, 40, 319
\bibitem[Chardonnet(2001)]{Char} Chardonnet, P. 2001, Proc. of the 27th ICRC 2001 (Hamburg, Germany),1764
\bibitem[Dolag et al.(2004)] {Dolag} Dolag, K., Grasso, D., Springer, V. \& Tkatchev, I. 2004, JKAS, 37, 427 
\bibitem[Furlanetto \& Loeb(2001)]{Furlanetto} Furlanetto, S. R., \& Loeb, A. 2001, \apj, 556, 619
\bibitem[Han(2001)]{Han} Han, J. L. 2001, preprint (astro-ph/0110319)
\bibitem[Kachelriess, Serpico \& Teshima(2005)]{kst} Kachelriess, M., Serpico, P. D., \& Teshima M. 2005, preprint (astro-ph/0510444)
\bibitem[Kim et al.(1989)]{Kim} Kim, K., Kronberg, P. P., Giovannini, G., \& Venturi, T. 1989, Nature, 341, 720
\bibitem[Krause(2003)]{krause} Krause, M. 2003, preprint (astro-ph/0304245)
\bibitem[Kronberg et al.(2001)]{Kronberg} Kronberg, P. P., Dufton, Q. W., Li, H., \& Colgate, S. A. 2001, \apj, 560, 178
\bibitem[Linsley et al.(1962)]{Linsley} Linsley J., Scarsi L., \& Rossi B. 1962, Suppl. J. Phys. Soc. Japan 17, 91
\bibitem[Longair(1997)]{Longair} Longair, M. S. 1997, High Energy Astrophysics, Vol. 2 (2nd Edition; Cambridge: Cambridge University Press)
\bibitem[Medina-Tanco(1998)]{Tanco} Medina Tanco, G. A. 1998, \apj, 505, L79 
\bibitem[Olinto(2005)]{Olinto1} Olinto, A.V. 2005, AIP Conf. Proc. 745, High Energy Gamma-Ray Astronomy, ed. Felix A. Aharonian, Heinz J. V\"olk, \& Dieter Horns (New York), 48O
\bibitem[O'Neill, Olinto \& Blasi(2001)] {ONeil} O'Neill, S., Blasi, P. \& Olinto, A.V. 2001, Proc. of 27th ICRC 5 (Hamburg, Germany), 1999
\bibitem[Paturel et al.(2005)]{Paturel} Paturel, G., Vauglin, I., Petit, C., Borsengerger, J., Epchtein, N., Fouqu\'e, P., \& Mamon, G. 2005, \aap, 430, 751
\bibitem[Prouza \& Smida(2003)]{Prouza} Prouza, M., \& Smida, R. 2003, \aap, 410, 1
\bibitem[Ptak et al.(2004)]{Ptak} Ptak, A., Terashima, Y., Ho, L. C., \& Quataert, E. 2004 \apj, 606, 173
\bibitem[Rachen \& Biermann(1993)]{rb} Rachen, J. P., \& Biermann, P. L. 1993, \aap, 272, 161

\bibitem[Schluter \& Biermann(1950)]{Biermann} Schluter, A., \& Biermann, L. 1950 Z. Naturforsch. 5A, 237
\bibitem[Sigl, Miniati \& Ensslin(2004)]{Sig} Sigl, G., Miniati, F. \& Ensslin, T. A. 2004, \prd, 70, 043007
\bibitem[Stanev(1997)]{Stanev} Stanev, T. 1997, \apj, 479, 290
\bibitem[Takeda et al.(1999)]{agasa} Takeda, M. \emph{et al.} 1999, \apj, 522, 225; see also Hayashida \emph{et al.} astro-ph/0008102.
\bibitem[Vall\'ee(2004)]{Vallee} Vall\'ee, J.P. 2004, New Astronomy Review, 48, 763
\bibitem[Vietri(1995)]{vietri} Vietri, M. 1995, \apj, 453, 883
\bibitem[Yoshiguchi, Nagataki \& Sato(2004)]{yoshi} Yoshiguchi, H., Nagataki, S. \& Sato, K. 2004, \apj, 614, 43.
\bibitem[Zwicky \& Karpowicz(1964)]{zwi} Zwicky, F. \& Karpowicz, M. 1964, \aj, 69, 759.
\end{thebibliography}
\end{document}